\documentclass[12pt]{article}
\usepackage{amssymb}
\usepackage{amsmath}
\usepackage{latexsym}
\usepackage{epsfig}

\newif{\ifcomentarios}
\comentariosfalse
\input{tcilatex}

\begin{document}

\author{Carla Goldman\thanks{%
Corresponding author} and Elisa T. Sena \\
Departamento de F\'{\i}sica Geral - Instituto de F\'{\i}sica \\
Universidade de S\~{a}o Paulo CP 66318\\
05315-970 S\~{a}o Paulo, Brazil.}
\title{The dynamics of cargo driven by molecular motors in the context of
asymmetric simple exclusion processes.}
\date{September 8, 2008}
\maketitle

\begin{abstract}
We consider the dynamics of cargo driven by a collection of interacting
molecular motors in the context of an asymmetric simple exclusion processes
(ASEP). The model is formulated to account for i) excluded volume
interactions, ii) the observed asymmetry of the stochastic movement of
individual motors and iii) interactions between motors and cargo. Items (i)
and (ii) form the basis of ASEP models and have already been considered in
the literature to study the behavior of motor density profile \cite%
{parmeggiani 03}. Item (iii) is new. It is introduced here as an attempt to
describe explicitly the dependence of cargo movement on the dynamics of
motors. The steady-state solutions of the model indicate that the system
undergoes a phase transition of condensation type as the motor density
varies. We study the consequences of this transition to the properties of
cargo velocity.

\end{abstract}

\section{Introduction}

Asymmetric simple exclusion processes (ASEP) are specially convenient for
describing general properties of dynamical systems consisting on a
collection of many-interacting particles in situations for which the
physicochemical characteristics of the components and thus the nature of
interactions, need not to be described in details \cite{derrida}, \cite%
{review evans}. Because of this, ASEP models have been used to study the
collective movement of molecular motors that happen at the microtubules
within cellular environment \cite{chowdhury 03}, \cite{parmeggiani 03}, \cite%
{parmeggiani 04}, \cite{lipowsky}. These are models that can be defined in
one-(spacial)-dimension and incorporate the asymmetry of the motion of
individual motors.

Since the initial proposal pointing out to ASEP as a possibility to describe
the collective dynamics of molecular motors \cite{chowdhury 03}, the general
interests are mainly focused on the properties of the system at different
boundary conditions that allow to make predictions on the stationary
currents or on the average motor velocities as function of external loading
forces \cite{campas 06}. Also, the effects of motor coordination onto the
process of pulling on fluid membranes \cite{campas 08} have been studied in
the literature in the context of discrete ASEP models with disorder \cite%
{krug}, \cite{evans disorder}. It shall be interesting then to investigate
whether this type of description can be extended to describe directly the 
\textit{movement of cargo} driven by motors to understand some of its
characteristics observed in experiments.

Cargo transport by motors happens at the cellular environment where simple
diffusion of vesicles or nutrients is severely limited by the presence of
innumerous structures inside the cytoplasm. Besides, it is known that virus
particles can take advantage of the existing transport mechanisms using
molecular motors as carriers to reach the interior of the cell \cite{virus}, 
\cite{kerstin}. Therefore, by studying the \ properties of a model that
leads to quantitative predictions on the movement of both motor and cargo
might be helpful to understand mechanisms to prevent cell infection and/or
to design more efficient drug carriers \cite{carriers}.

We have already worked on this problem to investigate the movement of cargo
in connection to the short-time behavior of the motor density profile
defined in the context of the continuum limit of an ASEP with periodic
boundary conditions \cite{Licht Gold}. Here, we present an alternative to
describe the long-time regime (steady-state) of the movement of cargo using
a discrete version of the model. The elementary dynamical processes, that
is, the processes at the level of individual particles moving on a defined
one-dimensional lattice with periodic boundary conditions are such to
account for two kinds of particles - the motors and the cargoes. Other
processes that take place on a lattice and involve two kinds of particles
with exclusion have already been explored in the literature to describe
diverse phenomena. The presence of the seminal \textit{second class particle}%
, for example, has been considered to study the microscopic properties of
shock fronts exhibited by the system due to density inhomogeneities\textbf{\ 
}\cite{ferrari}. This kind of model can also be used to study the properties
of the density profile in the presence of defects \cite{sasamoto}.

The important point we want to notice is that, up to the present, all kinds
of particles in the ASEP models presented in the literature are, in all
cases, provided with their own (intrinsic) dynamics. As a rule, an encounter
of a particle with another is just supposed to change their original
dynamics, or even to impede their movement (excluded-volume interactions),
by inducing modifications on the rates of the original stochastic movement.
This is the case, for example of the model cited above containing second
class particles. The elementary processes in this case are

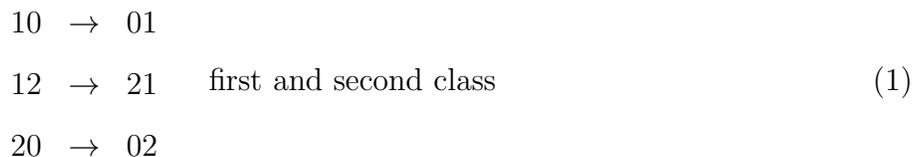
\begin{equation}
\begin{array}{ccc}
10 & \rightarrow & 01 \\ 
12 & \rightarrow & 21 \\ 
20 & \rightarrow & 02%
\end{array}%
\hspace{0.2in}\text{first and second class}  \label{second class}
\end{equation}%
that is, either particles of first $(1)$ and second $(2)$ classes can move
by their own if the target neighboring site is unoccupied $(0)$. They can
also move if interacting with other particles by interchanging occupancy
sites as in $12$ $\rightarrow 21$. This is also the case of another
ASEP-like model for which the particles with different dynamics simulate the
presence of cars $(1)$ and trucks $(2)$ on a same traffic road \cite{lee 97}%
, \textbf{\ }\cite{lee 99}. For this model, the direction of the intrinsic
movement of particle $(2)$ is opposed to that in the previous case:

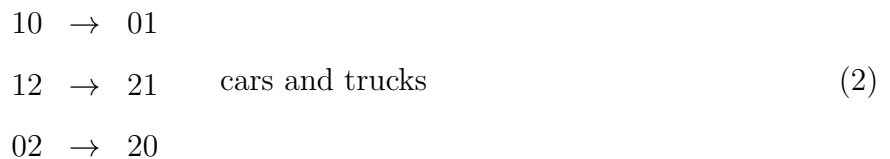
\begin{equation}
\begin{array}{ccc}
10 & \rightarrow & 01 \\ 
12 & \rightarrow & 21 \\ 
02 & \rightarrow & 20%
\end{array}%
\hspace{0.2in}\text{ cars and trucks}  \label{trucks}
\end{equation}

As we want to describe the movement of cargo driven by motors using an
ASEP-like model, we need first to think on possible ways cargo may use the
motors to move and up to what extent the presence of cargo can affect the
intrinsic dynamics of motors. The first idea that occurred to us is that the
elementary processes in this case should be such to attribute exclusively to
one kind of particles - the motors - the ability to move independently, i.e.
motors should be able to hop from site to site the only restriction being
excluded volume interactions with other motors. The other kind of particles
- the cargoes - should move only if \textit{assisted by motors}. Therefore,
the dynamics must, in addition, incorporate explicitly a form for the
motor/cargo interactions at this particle level. We want to examine how such
interactions affect the long-time average properties of both cargoes and
motors.

Based on the above considerations we present in Section 2 what we conceived
as a minimum model that is able to account explicitly for the dynamics of
both cargo and motors expected as a result from their mutual interactions.
This is presented as an ASEP model \ that incorporates a few common
characteristics of this biological system, but avoids details of motor/cargo
interactions. It turns out that this model is exactly solvable, that is, the
steady-state configurations can be determined. Using the matrix approach
proposed and developed by Derrida, Lebowitz, Evans, among others \cite%
{review evans}, \cite{derrida lebowitz}, these states are conveniently
represented as products of certain non-commuting matrices which can be used
to calculate the properties of interest as the average cargo velocity
discussed in Sec.3. The analytical results obtained in this way are used in
Section 4 to discuss the phenomenological consequences of the model.

\section{An ASEP model for motors and cargo}

There are a few proposals in the literature to characterize the origins and
the role of the components involved in the cargo/motor interactions at a
microscopic level \textbf{\ }\cite{kim 07}. Data suggest that such
interactions are mediated by certain proteins - dynactin is the most studied
- but apparently there is consensus about their short-range nature as a
general characteristic. In terms of the scales involved, this is equivalent
to say that such cargo/motor interactions happen by "direct contact" among
individual components. Moreover, although the experiments indicate that the
number of motors attached simultaneously to cargo may be large\textbf{\ }%
\cite{welte98}, there is limited information on the typical times each
motor, or group of motors, remains attached to cargo in the course of its
movement. Therefore, in building up our model, we avoid details of the
processes associated to such microscopic interactions and simply account for
these as stochastic processes. We suppose that the time scales associated to
the motor/cargo interactions are very short if compared with observation
time so the events that happen within such intervals need not to be
described as we treat the problem at a larger time scale.

The ASEP model considered here consists on a collection of $M$ motors, and $%
K $ cargoes, referred in the following as particles of type 1 and particles
of type 2 respectively, distributed among the $N$ sites of a one dimensional
cyclic lattice. Each particle occupies a single site and $N-M-K>0$ sites
remain empty (Fig.1). The total number of particles of each species is
conserved. We consider the case $K=1$ so that at the steady state, all
configurations of the system are equally likely (\footnote{%
In the presence of more than one particle of type 2 obeying the algebra (\ref%
{algebra}), the system loses ergodicity. In these more general cases, an
invariant measure should be assigned to each subspace of configurations that
preserve the number of empty sites between each pair of these particles.}).
Each site is identified by its position on the lattice $j=1,2,...N$ and
occupation at each site is specified by a corresponding site variable $%
\sigma _{j}$ that assumes integer values $0,1$ or $2$, if the site is empty (%
$0$), occupied by a motor ($1$), or occupied by cargo ($2$). A state $C$ of
the system is specified by the set $\{\sigma _{1}\sigma _{2}...$ $\sigma
_{N}\}.$ As usual, in this kind of description the stochastic dynamics is
defined through a Poissonian process taking place at the lattice such that
at each time interval $dt$, a pair of consecutive sites $i$ and $i+1$ is
selected at random and the system is updated depending on whether it is
possible to exchange particles between these two sites. We choose the
following possibilities 
\begin{equation}
\begin{array}{ccccc}
10 & \rightarrow & 01 & \text{with rate }k, & \hspace{0.1in}\text{%
probability }kdt \\ 
12 & \rightarrow & 21 & \text{with rate }w, & \hspace{0.1in}\text{%
probability }wdt \\ 
21 & \rightarrow & 12 & \text{with rate }p, & \hspace{0.1in}\text{%
probability }pdt%
\end{array}
\label{dynamics}
\end{equation}%
where the pair of sites $(i,i+1)$ is being represented by the values assumed
by the variables $(\sigma _{i},\sigma _{i+1}).$ According to these rules a
cargo is allowed to move only if \ "assisted"\ by a motor at a neighbor
site. We see this as a possibility to describe the fact that the movement of
cargo is conditioned to that of motors, as observed in real systems. At this
level of description one must consider that if motors affect the movement of
cargo then cargo must have influence on the movement of motors. In the
present model this is incorporated into the dynamics (\ref{dynamics}) both
explicitly, by the processes that involve cargo/motor exchanging positions
in both directions and implicitly, by modifying the motors hopping rates
that assume distinct values depending whether a jump occurs towards an empty
site or by interchanging position with particle 2.

Here, we are concerned with the kinematics of the cargo at long-time
regimes. For this, we use the matrix-approach \cite{derrida},\textbf{\ }\cite%
{review evans}\textbf{\ } to represent any configurations of the system of $%
N $ sites by a product of $N$ matrices, so that the probability of
occurrence of a particular configuration $C$ is given by%
\begin{equation}
P_{N,M}(C)=\dfrac{1}{Z_{N,M}}Tr\dprod\limits_{i=1}^{N}(\delta _{\sigma
_{i},1}D+\delta _{\sigma _{i},2}A+\delta _{\sigma _{i},0}E)
\label{prob do estado}
\end{equation}%
where the Kronecker delta symbols select the correct occupancy and 
\begin{equation}
Z_{N,M}=\sum_{\{\sigma _{i}\}}Tr\dprod\limits_{i=1}^{N}(\delta _{\sigma
_{i},1}D+\delta _{\sigma _{i},2}A+\delta _{\sigma _{i},0}E)  \label{norma}
\end{equation}%
is the normalization. The sum extends over all allowed configurations for
which $\sum_{i}^{N}\delta _{\sigma _{i},1}=M$ and $\sum_{i}^{N}\delta
_{\sigma _{i},2}=K=1$. In the product above, site $i$ is represented by
matrix $D$ if it is occupied\ by a motor (particle 1), by matrix $A$ if it
is occupied by the cargo (particle 2) or by matrix $E$ if it is empty
(not-occupied).

In the stationary state, the probabilities $P_{N,M}(C)$ for all
configurations $C$ satisfy the condition \cite{derrida lebowitz}:%
\begin{equation}
\sum_{C^{\prime }}P_{N,M}(C^{\prime })W(C^{\prime }\rightarrow
C)-P_{N,M}(C)W(C\rightarrow C^{\prime })=0  \label{mestra}
\end{equation}%
where $W(C^{\prime }\rightarrow C)$ is the rate at which the exchange of
particles occur between neighboring sites so that all nonzero terms in the
above sum are those for which configurations $C$ and $C^{\prime }$ differ
from each other at most by the occupancy of a pair of sites.

As an example, we consider for $N=4$ , $K=1$ , $M=2$ the following
configuration $C=1201$. There is just one way to leave this configuration
that is through the process: $12\rightarrow 21$ with $W(C\rightarrow
C^{\prime })=w.$ On the other hand, there are two ways to reach this
configuration, either by exchanging particle positions (i) by the process $%
10\rightarrow 01$ in configuration $C^{\prime }=1210$ with $W(C^{\prime
}\rightarrow C)=k$ or (ii) by the process $21\rightarrow 12$ in
configuration $C^{\prime }=2101,$ with $W(C^{\prime }\rightarrow C)=p.$ So,
in this case, equation (\ref{mestra}) reads:%
\begin{equation}
wTr(DAED)=kTr(DADE)+pTr(ADED)  \label{exemplo1}
\end{equation}

As a general rule, the main difficulty in using this method to determine the
probabilities $P_{N,M}(C)$ for each configuration $C$ is to find the
algebra, if any, that must be satisfied by the corresponding matrices of a
given ASEP model in order to satisfy condition (\ref{mestra}) for a given
dynamics as in (\ref{dynamics})$.$ In the present case, we conjecture that
if $D,A$ and $E$ are such that%
\begin{equation}
\begin{array}{c}
DA-xAD=E-D \\ 
DE=E \\ 
EA=E \\ 
EE=E%
\end{array}
\label{algebra}
\end{equation}%
for 
\begin{equation}
x=\frac{k+p}{w}  \label{x}
\end{equation}%
then, Eq. (\ref{mestra}) is satisfied. This can be tested using explicit
examples, as the one considered in (\ref{exemplo1}). Using (\ref{algebra})
to evaluate the traces, one can easily check that the identity holds
trivially. In the following, we study properties of this model that are of
interest for examining the consequences of the model regarding the
characteristics of cargo movement.

\bigskip

\section{ Average cargo velocity}

For $K=1$ i.e. just one cargo, in the presence of $M$ motors distributed
along a cyclic lattice of $N$ sites, we consider $N>M+1,$ to ensure that at
least one site in the system remains empty. In this case, the average cargo
velocity $<v>$ at steady state is expressed as%
\begin{equation}
<v>=\frac{1}{Z_{N,M}}\left( p\sum_{\{\sigma
_{i}\}}Tr\dprod\limits_{i=1}^{N-2}(\delta _{\sigma _{i},1}D+\delta _{\sigma
_{i},0}E)AD-w\sum_{\{\sigma _{i}\}}Tr\dprod\limits_{i=1}^{N-2}(\delta
_{\sigma _{i},1}D+\delta _{\sigma _{i},0}E)DA\right)  \label{v definicao 1}
\end{equation}%
where the sums in the numerator extends over all configurations of $M-1$
motors distributed among $N-2$ sites. The first sum in the RHS accounts for
all configurations in which the site at the immediate right of cargo is
occupied by a motor. The second sum accounts for all configurations in which
there is a motor at immediate left of cargo at a lattice position. Notice
that due to the invariance of the trace the normalization factor $Z_{N,M}$
can be written as 
\begin{equation}
Z_{N,M}=\sum_{\{\sigma _{i}\}}Tr\dprod\limits_{i=1}^{N-1}(\delta _{\sigma
_{i},1}D+\delta _{\sigma _{i},0}E)A  \label{norma 2}
\end{equation}%
where the sum extends over all configurations of $M$ motors distributed
among $N-1$ sites.

In order to make reference to the above traces over products of matrices, it
is convenient to introduce the functions $W_{\sigma _{j-1},\sigma
_{j,}\sigma _{j+1....}}$ to indicate the configurations having the n-uple $%
(j-1,j,j+1,...)$ fixed, \ the corresponding sites occupied by particles or
holes assigned by the variables $\sigma _{j-1},\sigma _{j},\sigma
_{j+1},.... $ Using these definitions, we write%
\begin{equation}
<v>=\frac{1}{Z_{N,M}}\left( p\sum_{\{\sigma _{i}\}}W_{21}-w\sum_{\{\sigma
_{i}\}}W_{12}\right)  \label{v definicao 2}
\end{equation}%
where%
\begin{equation}
Z_{N,M}=\sum_{\{\sigma _{i}\}}W_{21}+\sum_{\{\sigma _{i}\}}W_{20}
\label{norma 3}
\end{equation}%
for $W_{12}=\tprod\limits_{i=2}^{N-2}(\delta _{\sigma _{i},1}D+\delta
_{\sigma _{i},0}E)DA$ and analogous definitions for $W_{21},W_{20}$ and $%
W_{02}.$ Alternatively, due to the cyclic property of the trace, $Z_{N,M}$
can also be calculated from 
\begin{equation}
Z_{N,M}=\sum_{\{\sigma _{i}\}}W_{12}+\sum_{\{\sigma _{i}\}}W_{02}.
\label{norma 4}
\end{equation}%
These two expressions (\ref{norma 3}) and (\ref{norma 4}) are equivalent and
both will be used below, at convenience. Notice that we can rewrite the sum
over configurations in $W_{20}$ as 
\begin{equation}
\sum_{\{\sigma _{i}\}}W_{20}=\sum_{\{\sigma _{i}\}}W_{120}+\sum_{\{\sigma
_{i}\}}W_{020}  \label{w20}
\end{equation}%
where the first (second) sum on the RHS extends over all configurations of $%
M-1$ motors with the triplet $120$ fixed $(M$ motors with the triplet $020$
fixed) distributed among $N-3$ lattice sites. Making use of the
decompositions in (\ref{norma 3}) and (\ref{w20}), the average cargo
velocity (\ref{v definicao 2}) is expressed as 
\begin{equation}
<v>=p-\frac{1}{Z_{N,M}}\left\{ p\left( \sum_{\{\sigma
_{i}\}}W_{120}+\sum_{\{\sigma _{i}\}}W_{020}\right) +w\left( \sum_{\{\sigma
_{i}\}}W_{12}\right) \right\}  \label{v definicao 3}
\end{equation}%
We compute $Z_{N,M}$ as it is expressed in Eq. (\ref{norma 4}).

A convenient way to perform the calculations indicated above is to replace
the sums over site variables $\{\sigma _{i}\}$ by sums over blocks defined
by the integers $\{m_{i}\}$ and $\{q_{i}\}$ for $i=1,2...k.$ \ (see for
example, ref.\cite{lee 97}). In this representation, 
\begin{equation}
\sum_{\{\sigma
_{i}\}}W_{12}=\sum_{\{m_{i}\};\{q_{i}%
\}}tr(E^{q_{1}}D^{m_{1}}...E^{q_{k-1}}D^{m_{k-1}}E^{q_{k}}D^{m_{k}}A)
\label{sum 1 W12}
\end{equation}%
with $m_{k}\geq 1;$%
\begin{equation}
\sum_{\{\sigma
_{i}\}}W_{02}=\sum_{\{m_{i}\};\{q_{i}%
\}}tr(E^{q_{1}}D^{m_{1}}...E^{q_{k-1}}D^{m_{k-1}}E^{q_{k}}A)
\label{sum 1 W02}
\end{equation}%
with $q_{k}\geq 1;$%
\begin{equation}
\sum_{\{\sigma
_{i}\}}W_{120}=\sum_{\{m_{i}\};\{q_{i}%
\}}tr(E^{q_{1}}D^{m_{1}}...E^{q_{k-1}}D^{m_{k-1}}E^{q_{k}}D^{m_{k}}A)
\label{sum 1 W120}
\end{equation}%
with $q_{1}\geq 1$ and $m_{k}\geq 1;$%
\begin{equation}
\sum_{\{\sigma
_{i}\}}W_{020}=\sum_{\{m_{i}\};\{q_{i}%
\}}tr(E^{q_{1}}D^{m_{1}}...E^{q_{k-1}}D^{m_{k-1}}E^{q_{k}}A)
\label{sum 1 W020}
\end{equation}%
with $q_{1}\geq 1$ and $q_{k}\geq 1;$

From the algebra in (\ref{algebra}), it follows that $D^{m}AE=x^{m}AE.$ This
identity is needed in the evaluation of the above traces for general
configurations of the variables $\{\sigma _{i}\}$. The results are%
\begin{equation}
\begin{array}{ccc}
W_{12}\equiv & 
tr(E^{q_{1}}D^{m_{1}}...E^{q_{k-1}}D^{m_{k-1}}E^{q_{k}}D^{m_{k}}A)= & 
x^{m_{k}}tr(E) \\ 
W_{02}\equiv & tr(E^{q_{1}}D^{m_{1}}...E^{q_{k-1}}D^{m_{k-1}}E^{q_{k}}A)= & 
tr(E) \\ 
W_{120}\equiv & 
tr(E^{q_{1}}D^{m_{1}}...E^{q_{k-1}}D^{m_{k-1}}E^{q_{k}}D^{m_{k}}A)= & 
x^{m_{k}}tr(E) \\ 
W_{020}\equiv & tr(E^{q_{1}}D^{m_{1}}...E^{q_{k-1}}D^{m_{k-1}}E^{q_{k}}A)= & 
tr(E)%
\end{array}
\label{resultsW}
\end{equation}%
Notice that $W_{12}$ and $W_{120}$ are functions of the size $m_{k}$ of the
block, i.e. of the number of particles of type $1$ (motors) that precede
particle $2$ (cargo). Because of this, the evaluation of the respective sums
over $\{q_{i}\}$ and $\{m_{i}\}$ in configurations of the type $W_{12}$ (or $%
W_{120}$)$,$ excluding $m_{k},$ is equivalent to account for the number of
ways for distributing $M-m_{k}$ motors into $N-m_{k}-2$ (or into $N-m_{k}-3$%
) sites. The factor $2$ in the first case comes from the exclusion of two
sites from the total: one occupied by the cargo and another that must remain
empty to define the limits of the cluster of $m_{k}$ motors behind the
cargo. Then, 
\begin{equation}
S_{1}\equiv \sum_{\{\sigma _{i}\}}W_{12}=\sum_{m=1}^{M}\binom{N-m-2}{M-m}%
x^{m}tr(E)  \label{sum 2 W12}
\end{equation}%
and%
\begin{equation}
S_{2}\equiv \sum_{\{\sigma _{i}\}}W_{120}=\sum_{m=1}^{M}\binom{N-m-3}{M-m}%
x^{m}tr(E)  \label{sum 2 W120}
\end{equation}

\bigskip

$W_{02}$ and $W_{020}$ correspond to configurations that do not present
motors behind the cargo. In the sum over configurations of the type $W_{02}$
one must account for the number of ways to distribute $M$ motors into $N-2$
sites (from the total of $N$ sites, there must be excluded $2$, one to fix
the cargo and the other to fix an empty site). Then, 
\begin{equation}
\sum_{\{\sigma_{i}\}}W_{02}=\binom{N-2}{M}tr(E)  \label{sum 2 W02}
\end{equation}
and\textbf{\ }%
\begin{equation}
\sum_{\{\sigma_{i}\}}W_{020}=\binom{N-3}{M}tr(E)  \label{sum 2 W020}
\end{equation}

\bigskip

\bigskip

We now proceed by computing the sum over integer $m$ in (\ref{sum 2 W12})
and (\ref{sum 2 W120}).

\subsection{Aproximate expression for the average velocity of the cargo in
the limit of very large systems}

Our intention is to obtain an expression for the average velocity $<v>$ of \
cargo in the limit for which both the number of sites and the number of
motors (which are conserved by the dynamics) are taken very large, that is $%
N\rightarrow \infty $ and $M\rightarrow \infty .$ These limits are supposed
to be taken in such a way to ensure that the ratio between these two
quantities 
\begin{equation}
\lim_{\substack{ N\rightarrow \infty  \\ M\rightarrow \infty }}\frac{M}{N}%
=\rho  \label{limite termod}
\end{equation}%
converges to a defined density of motors $\rho $, such that $0<\rho <1.$
These same limits have already been considered in Ref. \cite{lee 97} to
calculate the average velocity of trucks in a related traffic problem. Here,
we proceed along the same lines sketched by these authors.

First, we use Stirling formula $N!\sim \sqrt{2\pi N}N^{N}\exp (-N)$ for very
large $N$, to approximate the combinatorial coefficients in (\ref{sum 2 W02}%
) and (\ref{sum 2 W020}). It results%
\begin{equation}
C_{02}\equiv \lim_{\substack{ N\rightarrow \infty  \\ M\rightarrow \infty }}%
\binom{N-2}{M}\sim \lim_{N\rightarrow \infty }\frac{(1-\rho )^{2}\exp
[-N(\rho \ln \rho +(1-\rho )\ln (1-\rho ))]}{\sqrt{2\pi N\rho (1-\rho )}}
\label{C02}
\end{equation}%
and%
\begin{equation}
C_{020}\equiv \lim_{\substack{ N\rightarrow \infty  \\ M\rightarrow \infty }}%
\binom{N-3}{M}\sim \lim_{N\rightarrow \infty }\frac{(1-\rho )^{3}\exp
[-N(\rho \ln \rho +(1-\rho )\ln (1-\rho ))]}{\sqrt{2\pi N\rho (1-\rho )}}
\label{C020}
\end{equation}%
In order to evaluate the sums $S_{1}$ and $S_{2}$ in (\ref{sum 2 W12}) and (%
\ref{sum 2 W120}), we follow the procedure used in Ref.\cite{marchetti}.
There, sums involving factorials of this kind are approximated by integrals
and the asymptotic regimes are obtained using Laplace's method \cite{murray}%
. Considering then the limit of very large systems and defining $\ z=m/N$,
we calculate 
\begin{equation}
\lim_{\substack{ N\rightarrow \infty  \\ M\rightarrow \infty }}S_{1}\simeq
\lim_{N\rightarrow \infty }\frac{\sqrt{N}(1-\rho )^{2}}{\sqrt{2\pi (1-\rho )}%
}\dint\limits_{0}^{\rho }\left( \frac{1-z}{\rho -z}\right) ^{1/2}\frac{%
e^{Nf(z)}}{(1-z)^{2}}dz  \label{S1 W12 limit}
\end{equation}%
where we have defined the function $f(z)$ of a single variable $z$ as 
\begin{equation}
f(z)=(1-z)\ln (1-z)-(\rho -z)\ln (\rho -z)-(1-\rho )\ln (1-\rho )+z\ln x
\label{f(z)}
\end{equation}%
and used the fact that in the specified limit, the sum in $m$ converges to
the integral as $\frac{1}{N}\sum\limits_{m}^{M}\rightarrow
\tint\limits_{0}^{\rho }dz$.

Now observe that $f(z)$ has a maximum at%
\begin{equation}
z_{\max }=\frac{1-x\rho }{1-x},  \label{zmax}
\end{equation}%
so in order to apply Laplace's method in the present case, one must
distinguish between two possibilities, namely

\begin{itemize}
\item if $x\rho \geq 1$ ($x>1$) then, $0\leq z_{\max }\leq $ $\rho $ i.e. $%
z_{\max }$ belongs to the integration interval. \ In this case, Laplace's
method gives 
\begin{equation}
\lim_{\substack{ N\rightarrow \infty  \\ M\rightarrow \infty }}S_{1}\sim
\left( \frac{x-1}{x}\right) \exp [-N(-\ln x+(1-\rho )\ln (x-1))]
\label{S1Laplace 1}
\end{equation}%
or

\item if $x\rho \leq 1$ either for $x>1$ or $x<1$, then $z_{\max }<0$.
Therefore, in this case $z_{\max }$ does not belong to the integration
interval. Since $f(z)$ is a monotone decreasing function of $z,$ the
integral is dominated by the value of the integrand at $z=0$, and
application of Laplace's method results 
\begin{equation}
\lim_{\substack{ N\rightarrow \infty  \\ M\rightarrow \infty }}S_{1}\sim -%
\frac{(1-\rho )^{2}}{\sqrt{2\pi N\rho (1-\rho )}}\frac{1}{\ln (\rho x)}\exp
[-N(\rho \ln \rho +(1-\rho )\ln (1-\rho ))]  \label{S1Laplace 2}
\end{equation}
\end{itemize}

Analogously, the asymptotic behavior of the sum $S_{2}$ (\ref{sum 2 W120})
must be analyzed according to the range of $x\rho$:

\begin{itemize}
\item if $x\rho \geq 1$ $(x>1),$ then 
\begin{equation}
\lim_{\substack{ N\rightarrow \infty  \\ M\rightarrow \infty }}S_{2}\sim
\left( \frac{x-1}{x}\right) ^{2}\exp [-N(-\ln x+(1-\rho )\ln (x-1))]
\label{S2  Laplace 1}
\end{equation}%
or

\item if $x\rho \leq 1,($either $x>1$ or $x<1),$ then 
\begin{equation}
\lim_{\substack{ N\rightarrow \infty  \\ M\rightarrow \infty }}S_{2}\sim -%
\frac{(1-\rho )^{3}}{\sqrt{2\pi N\rho (1-\rho )}}\frac{1}{\ln (\rho x)}\exp
[-N(\rho \ln \rho +(1-\rho )\ln (1-\rho ))]  \label{S2  Laplace 2}
\end{equation}
\end{itemize}

From these results, one concludes that if $x\rho \geq 1$ then $\ $both $%
S_{1} $ and $S_{2}$ are the dominant factors in the expression for $<v>$
both in the numerator and in the denominator. In this regime,%
\begin{equation}
\lim_{\substack{ N\rightarrow \infty  \\ M\rightarrow \infty }}<v>\sim \lim 
_{\substack{ N\rightarrow \infty  \\ M\rightarrow \infty }}\left[ p-\frac{1}{%
S_{1}}(pS_{2}+wS_{1})\right] \sim -\frac{kw}{k+p}\hspace{0.1in}\hspace{0.1in}%
\text{for}\hspace{0.25in}x\rho \geq 1  \label{v maior xrho}
\end{equation}%
On the other hand, if $x\rho \leq 1$, all factors in the expression (\ref{v
definicao 3}) for $<v>$ are of the same order of magnitude and then,%
\begin{equation}
\lim_{\substack{ N\rightarrow \infty  \\ M\rightarrow \infty }}<v>\sim \lim 
_{\substack{ N\rightarrow \infty  \\ M\rightarrow \infty }}\left\{ p-\frac{1%
}{\left( S_{1}+C_{02}\right) }\left[ p\left( S_{2}+C_{020}\right) +wS_{1}%
\right] \right\} \sim p\rho +\frac{w}{\ln (\rho x)-1}\hspace{0.1in}\hspace{%
0.1in}\text{for}\hspace{0.25in}x\rho \leq 1  \label{v menor xrho}
\end{equation}%
The consequences of these results to the phenomenology of cargo movement
will be analyzed in the next section.

\section{Discussion and concluding remarks}

The aim of the present work is to study the consequences of introducing
cargo into certain lattice models where there is also present a set of
biased molecular motors interacting through excluded-volume interactions. We
consider an ASEP-like model specially designed to account for both kinds of
particles. Therefore, the model includes an assumption about the stochastic
nature of the movement of cargo and its dependence on the dynamics of
motors. To our knowledge, this is the first attempt to include in the same
framework, at the particle level, the effects on the movement of motors due
to the presence of cargo and vice-versa.

We look for the probabilities associated to the configurations of the system
at the stationary state which are represented by products of certain
noncomuting matrices \cite{derrida}. Using this representation, we were able
to make quantitative predictions on the average properties that characterize
the movement of cargo. We focus on the computation of the average velocity
of cargo $<v>$ whose behavior predicted by the model suggests that the
system displays a phase transition under variation of the parameters. Fig. 2
shows $<v>$ plotted according to the results in Eqs. (\ref{v maior xrho})
and (\ref{v menor xrho}) as a function of $\rho ,$ at different values of $\
p$ for fixed $w$ and $k$. Observe that for sufficiently high values of $p$
the function $<v>$ displays a change in its behavior at values of $\rho $
for which $x\rho =1,$ as $<v>$ becomes independent of$\ $\ $\rho $.

In order to interpret these results, it shall be easier first to discuss on
the kind of movement one would expect for cargo in the context of the
considered ASEP. The mechanisms in (\ref{dynamics}) that define its
elementary movements within each unit interval of time correspond to those
of exchanging positions with a neighbor motor. The assigned hopping rates
are such to promote, at a first moment, an accumulation of motors at one
side (at the left side) of the cargo. Then cargo would be able to move
backwards by exchanging position with these accumulated motors. By doing
this, the motors end up transposed to the cargo's front. Because motors are
assigned with an intrinsic dynamics - they move preferentially to the right
- these motors at cargo's front will tend to disperse. Since the vesicle
depends on such clusters of motors to develop a measurable velocity, it ends
up moving mostly due to the motors accumulated at its back. \ \ \ \ \ \ \ \ 

The dependence of $<v>$ on $\rho $ in Fig.2 confirms these expectations
showing that $<v>$ assumes only negative values, at all ranges of
parameters. One could expect, in principle, that at high values of $\rho $
there would be a balance between a tendency for maintenance of motors in
front of the cargo, as excluded-volume become more important and eventually
would be responsible for expressive motor accumulation at cargo's front. So,
in principle, one could think that for sufficient high values of $p$ there
would be a chance for the cargo to develop a macroscopic movement towards
the plus end of the microtubule, that is to the forward direction as well, $%
<v>$ eventually displaying positive values\ for $\rho >x^{-1}.$ According to
the results, however, this does not happen either. The behavior $\ $of $\
<v> $ predicted for high values of $\rho $ can be\textbf{\ }understood by
recognizing the formation of an \ "infinite" (macroscopic) cluster of motors
behind the cargo with which it can always exchange positions. The formation
of such infinite cluster\ would be a consequence of the phase transition (of
condensation type) predicted for this system. The average velocity in this
region of density becomes constant probably due to the uniformity of motor
distribution along this cluster.

We can then summarize these results by saying \ that the single cargo in
this system of many motors develops, a backwards movement at any value of
the hopping rates $k,p$ and $w$, or density $\rho .$ The magnitude of such
velocity, however, is highly dependent on $\rho $ and becomes constant at
such values of $\rho $ greater than a critical value $\rho _{c}=x^{-1}.$ It
would be interesting then to test these predictions using data from
experiment \textit{in vivo} by monitoring the behavior of the motor density
as cargo moves.

Actually, data from Drosophila embryos \cite{welte98}, \cite{gross}\textbf{\ 
}show that cargo velocity presents distinct behaviors depending on the stage
of embryo development. It remains to investigate whether these changes could
be associated to corresponding changes in the density of motors available at
each of these stages. To our knowledge, there is limited information about
the possible changes on the motor distribution along the microtubules due to
the movement of cargo. The present study suggests that any investigation in
this direction might be relevant to find ways to control cargo movement.

\bigskip

\textbf{Acknowledgments}

\bigskip

We acknowledge the financial support from Funda\c{c}\~{a}o de Amparo \`{a}
Pesquisa do Estado de S\~{a}o Paulo (FAPESP).

\newpage

\textbf{Figure Caption}

\vspace{0.3in}

\begin{itemize}
\item Fig.1 - A configuration of the one-dimensional (discrete) ASEP model
for interacting motors (gray) and cargo (black). Each particle, occupies a
single site at each instant of time. The non-occupied (empty) sites are
represented by line segments. The random processes are such that each motor
is allowed to hop at rate $k$ to its nearest neighbor on the right if it is
empty. The motors and cargo can exchange places at a rate $p$ (motor jumping
to nearest neighbor at the right ) or $q$ (motor jumping to the nearest
neighbor at the left).

\item Fig.2 - Average velocity of cargo as a function of motor density $\rho 
$, at various values of parameter $p$, for fixed $w=3$ and $k=1,$ as
indicated. The predicted phase transition is illustrated by the change in
the behavior of $<v>$ that assume a constant value for $\rho >\rho _{c}=1/x$.
\end{itemize}

\end{document}